\documentclass[iop]{emulateapj}
\usepackage{graphicx}
\usepackage{enumerate}
\usepackage{amssymb, amsmath}
\usepackage{natbib}
\usepackage{color}
\usepackage{ulem}
\usepackage{dblfnote}
\usepackage{appendix}

\def\pasa{PASA}
\def\procspie{Proc.~SPIE}

\renewcommand{\thefootnote}{\alph{footnote}}

\newcommand{\tokyo}{1}
\newcommand{\ichinoseki}{2}
\newcommand{\abc}{3}
\newcommand{\naoj}{4}
\newcommand{\tokyoeps}{5}
\newcommand{\ias}{6}

\begin{document}

\author{Taichi Uyama$^\tokyo$}
\author{Takayuki Tanigawa$^\ichinoseki$}
\author{Jun Hashimoto$^\abc$}
\author{Motohide Tamura$^{\tokyo,\abc,\naoj}$}
\author{Yuhiko Aoyama$^\tokyoeps$}
\author{Timothy D.Brandt$^\ias$}
\author{Masato Ishizuka$^\tokyo$}

\footnotetext[\tokyo]{Department of Astronomy, The University of Tokyo, 7-3-1, Hongo, Bunkyo-ku, Tokyo 113-0033, Japan}
\footnotetext[\ichinoseki]{National Institute of Technology, Ichinoseki College, Hagisho Takanashi, Ichinoseki 021-8511, Japan}
\footnotetext[\abc]{Astrobiology Center of NINS, 2-21-1, Osawa, Mitaka, Tokyo 181-8588, Japan}
\footnotetext[\naoj]{National Astronomical Observatory of Japan, 2-21-1, Osawa, Mitaka, Tokyo 181-8588, Japan}
\footnotetext[\tokyoeps]{Department of Earth and Planetary Science, The University of Tokyo, 7-3-1, Hongo, Bunkyo-ku, Tokyo 113-0033, Japan}
\footnotetext[\ias]{Astrophysics Department, Institute for Advanced Study, Princeton, NJ, USA}

\title{Constraining accretion signatures of exoplanets in the TW Hya transitional disk}

\begin{abstract}
We present a near-infrared direct imaging search for accretion signatures of possible protoplanets around the young stellar object (YSO) TW Hya, 
a multi-ring disk exhibiting evidence of planet formation.
The Pa$\beta$ line (1.282 $\mu$m) is an indication of accretion onto a protoplanet, and its intensity is much higher  than that of blackbody radiation from the protoplanet.
We focused on the Pa$\beta$ line and performed Keck/OSIRIS spectroscopic observations.
Although spectral differential imaging (SDI) reduction detected no accretion signatures, the results of the present study allowed us to set 5$\sigma$ detection limits for Pa$\beta$ emission of $5.8\times10^{-18}$ and $1.5\times10^{-18}$ erg/s/cm$^2$ at 0\farcs4 and 1\farcs6, respectively. We considered the mass of potential planets using theoretical simulations of circumplanetary disks and hydrogen emission. The resulting masses were $1.45\pm 0.04$ M$_{\rm J}$ and $2.29 ^{+0.03}_{-0.04}$ M$_{\rm J}$ at 25 and 95 AU, respectively, which agree with the detection limits obtained from previous broadband imaging. The detection limits should allow the identification of protoplanets as small as $\sim$1 M$_{\rm J}$, which may assist in direct imaging searches around faint YSOs for which extreme adaptive optics instruments are unavailable.

\end{abstract}

\section{Introduction} 
Young stellar objects (YSOs) often have protoplanetary disks in which planets are formed. If a protoplanet exists, it should be apparent by the effect it has on the disk.
YSOs with protoplanetary disks have infrared (IR) excesses in their spectral energy distribution (SED). Some show far-infrared (FIR) excesses but little excess in the mid-infrared (MIR) region. Objects with such characteristic SEDs are called ``transitional disks'', and may suggest inner gaps in the disk and ongoing planet formation \citep[][]{Marsh1992,Marsh1993}.
Protoplanetary disks with intriguing features, such as spiral arms, multiple rings, and large gaps, have been discovered based on IR and millimeter/sub-millimeter wavelength studies \citep[e.g.][]{Andrews2011,Hashimoto2011,Hashimoto2012,Mayama2012,Muto2012,ALMA2015}. Theoretical calculations can estimate the characteristics of potential protoplanets based on the derived parameters for these disks  \citep[][]{Dong2015}.
Adaptive optics (AO) based observations have detected substellar-mass companion candidates \citep[e.g., HD 169142 and LkCa 15;][]{Reggiani2014,Sallum2015} within gaps in protoplanetary disks. Compared to the number of gapped or asymmetric disk discoveries, however, few companion candidates have been detected around YSOs, which may be due to the contrast being too low to detect faint objects at small separations.
The typical detection limits for previous YSO surveys are given in \cite{Uyama2017}.

Extreme AO (ExAO) observations such as VLT/SPHERE \citep[][]{Beuzit2006}, Gemini/GPI \citep[][]{Macintosh2006}, and Subaru/SCExAO \citep[][]{Guyon2010} enable us to overcome the difficulty due to low contrast. However, these instruments are limited to observations of brighter target ($R\leq$11--12 mag) YSOs.

Here, we focus on the use of classical AO techniques to detect accretion signatures of protoplanets based on their hydrogen emission spectra. In the process of planet formation, accretion shocks excite the surrounding hydrogen. When the excited hydrogen returns to lower energy state, emission lines such as H$\alpha$ are produced \citep[][]{Calvet1998}. 
The intensity of hydrogen emission lines is much higher than that of blackbody radiation from the protoplanet \citep[e.g.][]{Bowler2014,Zhou2014}.
For an isolated stellar-mass object case, hydrogen emission is produced in shock-heated gas from the circumstellar disk to the stellar surface,
whereas gas accretion onto a planetary-mass companion leads to another strong shock structure associated with flow from the circumstellar disk to the circumplanetary disk \citep[e.g.][]{Tanigawa2012}. 
Considering this circumplanetary mechanism, the luminosity associated with hydrogen emission in the circumplanetary disk can be estimated as a function of the planetary mass and the gas density in the circumstellar disk \citep[][]{Aoyama2017P}.

In order to explore potential line emission from accreting protoplanets, we first observed TW Hya, one of the nearest YSOs. The stellar parameters for this YSO are listed in Table \ref{stellar parameter}. The SED for TW Hya indicates a transitional disk with active accretion \citep[][]{Calvet2002,Goto2012,Menu2014}.
Various observations have led to the conclusion that its disk contains multiple gaps and that the most likely locations for ongoing planet formation are at 0\farcs4 and 1\farcs6, where the disk's surface and midplane have dents \citep[][]{Debes2013,Menu2014,Akiyama2015,Tsukagoshi2016,Boekel2016}. There have been no reports of companion candidates \citep[e.g.][]{Boekel2016}. 

Section \ref{sec: Observations and Results} describes the observations and results. In Section \ref{sec: Discussions}, we set detection limits for low-mass objects. This work mainly focuses on accretion signatures based on the emission line intensity. 
Conventionally, emission line intensity is converted into accretion luminosity. However, if there is only one observable, there is a degeneracy between mass and mass accretion rate. Therefore, we present an additional method for interpreting the obtained data and evaluating the mass of protoplanets.  
Finally, we summarize the results of this study in Section \ref{sec: Summary}.

\begin{table}[t]  \centering
  \caption{Stellar Parameters for TW Hya}
  \begin{tabular}{ccc}\\ \hline\hline
  parameters & Value & Ref. \\ \hline
  Sp type & M2 & a \\
  Mass [$M_{\odot}$] & 0.55$\pm$0.15 & a \\
  Age [Myr] & 7--10 & b \\
  Distance [pc] & 59$\pm$1 & c \\
  Av & 0.0 & d \\
  $R$ mag & 10.4 & e \\
  $J$ mag & 8.22 & f \\
  $J$-band flux [erg/s/cm$^{2}$/$\mu$m] & 1.60$\times10^{-9}$ & f \\ \hline
  \end{tabular}
  \label{stellar parameter}
  \tablecomments{a: \cite{Debes2013}, b: \cite{Akiyama2015}, c: \cite{Gaia2016}, d: \cite{Herczeg2004}, e: UCAC4 catalogue \citep[][]{Zacharias2012} f: 2MASS catalogue \citep[][]{Cutri2003}}
\end{table}

\section{Observations and Results}\label{sec: Observations and Results}
\subsection{Keck/OSIRIS}

\begin{figure}
    \centering
    \includegraphics[scale=0.33]{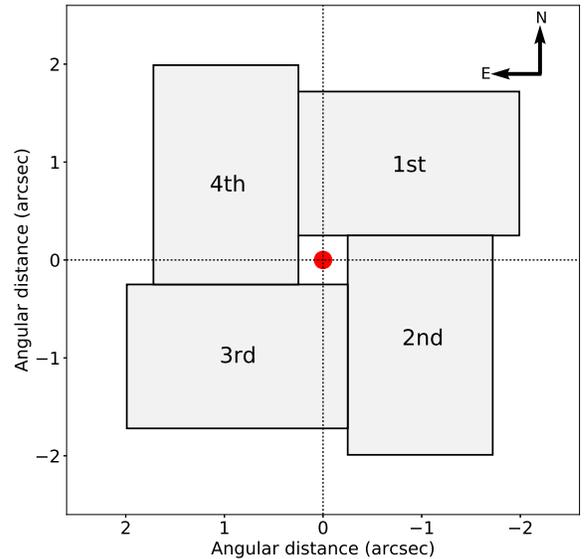}
    \caption{FOV arrangement for observations. Each rectangle represents a separate FOV. The red circle indicates the central star. The vertical and horizontal axes represent the angular distance from the central star. These settings avoid the central star together with an inner square of 0.5\arcsec$\times$0.5\arcsec, which allows sufficient exposure time without saturation.}
    \label{schematic}
\end{figure}

We performed Keck/OSIRIS observations in the Jn2-band (1.228--1.289 $\mu$m) with a 0\farcs035 plate scale, which corresponds to a 1.47\arcsec $\times$ 2.24\arcsec\ field of view (FOV) and a spectral resolution of $R\sim$3800. 
Pa$\beta$ (1.282 $\mu$m) and Br$\gamma$ (2.166 $\mu$m) lines are available for OSIRIS. The AO performance at Br$\gamma$ is better than that at Pa$\beta$ \citep[][]{vanDam2004}, while expected line intensity of Br$\gamma$ is much smaller than that of Pa$\beta$. We compared the feasibility of detecting these lines with OSIRIS and finally selected more favorable case of the Pa$\beta$ line to search for accretion signatures.
We set four FOVs around the central star to provide sufficient exposure time to avoid saturation of the stellar point-spread function (PSF), to perform a deep imaging survey. 
A schematic of the observation region is shown in Figure \ref{schematic}. We separated the FOVs 0.25\arcsec\ from the central star and explored the disk from $\sim$20 AU to $\sim$100 AU.

The observation was conducted on 2016 April 27. The exposure time for each FOV was 15 minutes $\times$ 3 and the total observation time was 180 minutes. We also observed unsaturated frames of the central star for telluric correction, photometric standard, and PSF reference. {We note that the OSIRIS enables to use the unsaturated frames as calibration reference regardless of its variability. Details are described in the following paragraph.}

After the first reduction using the OSIRIS pipeline, which corrects dark, flat, distortion, wavelengths solution, and cosmic rays, we removed telluric contributions from the data cube. In the telluric correction process the pipeline usually removes hydrogen, which means Pa$\beta$ in our data cube is not affected by telluric correction.
To achieve high contrast, we used a spectral differential imaging technique \citep[SDI;][]{Smith1987}.
We selected seven channels around the Pa$\beta$ wavelength as a science channel with a width of 1.05 nm and other 24 channels between 1.278 and 1.285 $\mu$m as continuum (reference) channels. We avoided the OH airglow region \citep[][]{Maihara1993} when selecting reference channels. The airglow sky lines remain in the data cube because they are used for wavelength references in the pipeline. 
We made reference PSFs for each FOV using linear combinations of the reference frames based on equations (2)--(4) in \cite{Artigau2008} to determine the coefficients of the reference images. In this process, we divided the FOV into 12 annular regions in a similar manner to the locally optimized combination of images (LOCI) method \citep[][]{Lafreniere2007}. 

\begin{figure}
    \centering
    \includegraphics[scale=0.35]{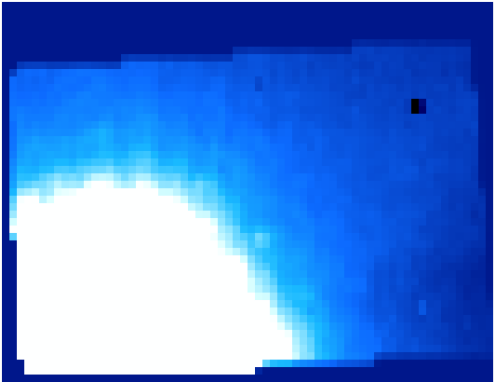}
    \caption{Pa$\beta$ image of the first FOV. The image is aligned with that in Figure \ref{schematic}. The black square near the upper-right vertex is a badpixel cluster.}
    \label{Pabeta}
\end{figure}

\begin{figure}
    \centering
    \includegraphics[scale=0.35]{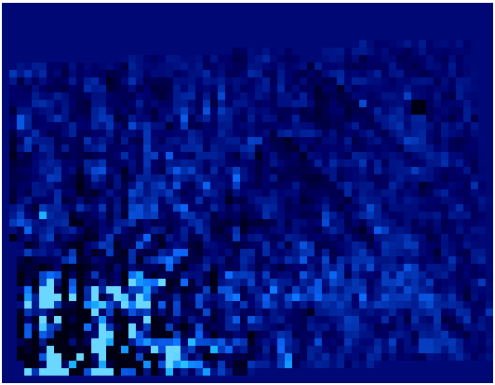}
    \caption{SDI-reduced image of the first FOV. The alignment is the same as in Figure \ref{Pabeta}.}
    \label{SDI-reduced}
\end{figure}

Following SDI data reduction, no point sources were detected in any of the FOVs. 
Figure \ref{Pabeta} shows the Pa$\beta$ image that was produced by combining the seven channels of the science frames, and Figure \ref{SDI-reduced} shows the final SDI-reduced image for the first FOV.
In the Pa$\beta$ image, because of low AO performance, a stellar halo is evident near the central star where there is expected to be remnants of subtraction in the SDI-reduced image due to Poisson noise.
We then normalized all of the images by dividing by their integration time, and convolved them with an aperture with a radius equal to half the FWHM of the central star. The FWHM was measured in the unsaturated frames to be $\sim$60 mas.
We calculated the standard deviation for each FOV and produced a radial noise profile. We masked badpixels in this process and they do not affect on calculating the noise.
The typical noise was compared to photometric results for the central star. The photomectic reference was made by combining the same continuum channels as the SDI reduction. We finally obtained 5$\sigma$ detection limits for the Pa$\beta$ flux
of $5.8\times10^{-18}$ and $1.5\times10^{-18}$ erg/s/cm$^2$ at 0\farcs4 and 1\farcs6, respectively. 


\subsection{IRSF/SIRIUS}
YSOs are variable due to stellar factors such as starspots, accretion, disks, and envelopes \citep[][]{Herbst1994,Wolk2013}.
While previous catalogues such as 2MASS \citep[][]{Cutri2003} contain photometric results for this YSO in the J-band, it is important to evaluate the variability of the central star in order to determine the detection limit for potential protoplanets. We used the J-band luminosity of TW Hya as a photometric reference. Investigating the variability helps in setting the error in the detection limit.

Since its luminosity varies over a period of a few days \citep[e.g][]{Rucinski2008,Siwak2014}, we observed TW Hya with IRSF/SIRIUS, a 1.4-m telescope in South Africa, using the defocus mode, on January 28 and 30, 2017 to examine its $J$-band variability.
After dark subtraction and flat fielding we conducted photometry. We found that the variability during the observations was less than 5\%, which agrees with the standard deviation in 2MASS, Tycho, UCAC 2, and other catalogues \citep[$\sim$7\%;][]{Cutri2003, Zacharias2004,Ofek2008,Tokovinin2012,Kunder2017}.
Since the detection limits are calculated based on the J-band flux of TW Hya, we define the error in the detection limit as 5\%.

\section{Discussion}\label{sec: Discussions}
We investigate the presence of potential protoplanets within TW Hya's disk in two different ways.
Section \ref{sec: Empirical Relationship} presents a conventional approach that assumes mass accretion from circumstellar disk onto the central star and ignores the inability to distinguish between the mass and mass accretion rate. Section \ref{sec: Simulation} proposes a new method for interpreting the detection limit by considering circumplanetary disk mechanisms. 
We focus on the two notable gaps at 0\farcs4 and 1\farcs6. We assume that these locations represent $\sim$25 AU and $\sim$95 AU using the distance data (59$\pm1$ pc) published by \cite{Gaia2016}, whereas previous studies have assumed that the gaps are at $\sim$20 AU and $\sim$80 AU based on data published by \cite{Van2007}.
We assume that extinction of the Pa$\beta$ line by the circumstellar disk is negligible. TW Hya's disk is almost face-on \citep[][]{Qi2004,Qi2008}. The vertical direction from the jovian-mass protoplanet can be estimated locally clear for Pa$\beta$ \citep[][]{Tanigawa2012,Aoyama2017P}. 

\subsection{Using the Empirical Relationship}
\label{sec: Empirical Relationship}

The Pa$\beta$ luminosity of YSOs has been reported to have an empirical correlation with the accretion luminosity \citep[][]{Natta2004,Natta2006} for objects down to $\sim$10 M$_{\rm J}$ \citep[e.g. GSC 06214-00210 b and FW Tau b;][]{Bowler2011,Bowler2014}.
The relationship is given by
\begin{eqnarray}
\log(\frac{L_{\rm acc}}{L_\odot}) = 1.36 \times \log(\frac{L_{\rm Pa\beta}}{L_\odot}) + 4,
\label{empirical relation}
\end{eqnarray}
where $L_{\rm acc}$ is the accretion luminosity and $L_{\rm Pa\beta}$ is the Pa$\beta$ luminosity. In this discussion, we assume that the empirical relationship is valid for objects as small as $\sim$1 M$_{\rm J}$.

Since we did not detect any accretion signals, constraining the mass of planets requires other assumptions for the accretion rate and radius. We assume an excessive radius and modest accretion rate because the accretion luminosity ($L_{\rm acc} = GM\dot{M}/R$, where $R$, $M$, and $\dot{M}$ are the radius, mass, and accretion rate, respectively) increases in proportion to the accretion rate and in inverse proportion to the radius.
We define an upper limit of $R_{\rm p}=10\ {\rm R_ J}$.
\cite{Ayliffe2009} and \cite{Szulagyi2016} suggested with nominal simulations that an object with ${\rm R_p}=10\ {\rm R_J}$ has a peak temperature of $\sim$$10^4$ K, which causes spectral emission from hydrogen.
The accretion rate for TW Hya is $9.0\times10^{-10}$--$4.2\times10^{-9}$ $M_\odot$/yr \citep[][]{Dupree2012}. Considering ongoing planet formation, the accretion rate for protoplanets should be close to that for the star \citep[][]{Tanigawa2016}. 
We define a lower limit of $\dot{M_{\rm p}} = 10^{-11}$ $M_\odot$/yr, which is smaller than the theoretical accretion rate of $\dot{M_{\rm p}} \simeq10^{-10}$ $M_\odot$/yr for a $M_{\rm p}$$\sim$1 M$_{\rm J}$ object \citep[][]{Lissauer2009}.
With these assumptions and the stellar parameters in Table 1, the calculation results give 5$\sigma$ detection limits of  $1.1\times10^{-2}$ and $1.7\times10^{-3}$ M$_{\rm J}$ at each gap. We do not think that these values can really constrain the mass of the protoplanets. The detection limits are calculated by extrapolation of Equation (\ref{empirical relation}) and thus less than the lower limit of the relationship in \cite{Natta2004}. These values correspond to Earth-mass object, which should be qualitatively too low mass to accrete hydrogen and to emit Pa$\beta$.
Therefore, the mass of possible protoplanets are not actually constrained if we use Equation (\ref{empirical relation}) with only Pa$\beta$ detection limits.

\subsection{Based on a Recent Theoretical Study}
\label{sec: Simulation}
A recent study of one-dimensional shock dynamics with detailed
radiative transfer, including line emissions and chemical reactions, has yielded values for the Pa$\beta$ flux and, as a result of integrating the flux over a circumplanetary disk. Assuming the circumstellar disk structure, total Pa$\beta$ luminosity from the circumplanetary disk can also be estimated \citep[][]{Aoyama2017P}.

Assuming a sufficiently strong shock, the gas velocity, density, and composition govern post-shock characteristics such as the energy of hydrogen emission lines \citep[e.g.][]{Landau+Lifshitz1959}.
Gas accretion in a circumstellar disk is governed by viscous dissipation with turbulence \citep[e.g.][]{Alexander2006}.
On the other hand, gas transfer from the gravity field of a star to that of a planet is driven by shock \citep[][]{Tanigawa2002}. Hydrodynamic simulations show that a protoplanet actively grows by capturing the circumstellar disk gas \citep[e.g.][]{DAngelo2003,Ayliffe2009}.
The accreting gas moving toward the planet forms shock surfaces at the top of the circumplanetary disk \citep[][]{Tanigawa2012,Szulagyi2014}.
In the thin layer of the post-shock region, the temperature of the gas can be very high ($>10^4$ K), which gives rise to hydrogen emission lines due to electron transitions \citep[][]{Szulagyi2017}.
We can convert the accretion rate to the number density of the circumstellar disk with these assumptions and thus the line luminosity can be written as a function of the planetary mass and the gas density in the circumstellar disk \citep[][]{Aoyama2017P}.
By performing this simulation using a disk model and a formula for TW Hya \citep[][]{Gorti2011}, we can estimate the luminosity of a possible protoplanet down to 0.5 M$_{\rm J}$. Note that the planetary radius and gas viscosity, which would be important factors to determine the line luminosity in the traditional models, hardly affect the luminosity in our shock model. Detailed concepts of the simulations are explained in Appendix.

We found that the luminosity of the circumplanetary disk around a $1$-M$_{\rm J}$ planet at 25 AU from the central star can be as high as
$4.1^{+0.4}_{-0.6}\times 10^{23}$ 
erg, which corresponds to $9\pm1 \times 10^{-19}$ erg/cm$^2$. 
Considering the detection limits ($5.8\times 10^{-18}$erg/cm$^2$ at 25AU, $1.5\times 10^{-18}$erg/cm$^2$), the upper limit for the observation is $1.45\pm 0.04$ M$_{\rm J}$ 
at 25 AU for an actively growing protoplanet, and 
$2.29 ^{+0.03}_{-0.04}$ M$_{\rm J}$ 
at 95 AU.
Although the detection limit for the inner gap is larger than that for the outer gap, the simulations can constrain the mass of a potential planet more strictly in the inner region because the gas density decreases with increasing radial distance from the central star.

\subsection{Comparison with Previous Imaging Studies}
\label{sec: Comparison}
The present study was based on a search for signatures of accretion by protoplanets, whereas previous studies have explored protoplanets themselves or disk structures indicative of planet formation.
Direct imaging observations combined with classical AO instruments gave typical 5$\sigma$ detection limits of $\sim$16 M$_{\rm J}$ and $\sim$3 M$_{\rm J}$ at 0\farcs4 and 1\farcs6 \citep[Subaru/HiCIAO;][]{Uyama2017}. We also compared the detection limits of VLT/SPHERE observation \citep[][]{Boekel2016} to evolutionary models of low-mass objects \citep[e.g. COND03 model, BT-Settl model;][]{Baraffe2003,Allard2011}, which suggests that potential planets are expected to be smaller than 0.5 M$_{\rm J}$ in each gap.
Despite uncertainties due to the assumptions made in the present study, the derived mass is comparable to that obtained from angular differential imaging \citep[ADI;][]{Marois2006} studies. Our observations show that classical AO instruments combined with SDI can feasibly set constraints on the mass of potential planets up to $\sim$1 M$_{\rm J}$ around faint YSOs, for which ExAO instruments cannot be used. 
Although we assumed zero extinction from the circumstellar disk in our study, we note that direct imaging studies have not overcome an issue of this extinction, which means we allow the uncertainty of the extinction from the circumstellar disk.

Disk observations, using both the Atacama Large Millimeter/submillimeter Array  (ALMA)  and polarization differential imaging (PDI) with AO instruments, together with simulation results, showed that ongoing planet formation can carve gaps in the disk. The predicted mass of a potential planet in the TW Hya gaps is 0.03--0.5 M$_{\rm J}$ \citep[e.g.][]{Akiyama2015,Dong2017,Teague2017}.
We note that our discussions assume the circumplanetary disk around an object more massive than 0.5 M$_{\rm J}$ and are independent from disk morphology.
The simulation results in Section \ref{sec: Simulation} are based on the number density of hydrogen atoms in the disk determined by radio wavelength observations, and are independent of the disk morphology.

Our estimated detection limits agree with the results of previous exoplanet and disk explorations in TW Hya, which suggests that protoplanets, if any, are so small that hydrogen emission spectra are not produced.

\section{Summary}\label{sec: Summary}
Previous direct imaging explorations of YSOs have focused on thermal emission from exoplanets themselves. We focus on emission signatures of accretion by protoplanets within the circumstellar disk, which are expected to be more luminous than blackbody radiation associated with an exoplanet.

We used Keck/OSIRIS to observe TW Hya, one of the most well known YSOs, which has a multi-gapped disk, in order to explore Pa$\beta$ emissions.
Using several FOVs, we developed an unconventional method for avoiding saturation while still allowing a sufficient exposure time. 
Although no signals associated with accretion were identified, Pa$\beta$ detection limits of $5.8\times10^{-18}$ and $1.5\times10^{-18}$ erg/s/cm$^2$ at 0\farcs4 and 1\farcs6 were determined, thus providing the first constraints on the mass of potential protoplanets.
We estimated the mass of protoplanets using both an empirical relationship and a disk simulation, each based on different assumptions. The first method is conventional and is based on the empirical relationship between the accretion luminosity and the Pa$\beta$ flux, and cannot separately determine the mass and the mass accretion rate.  The simulation assumes an active circumplanetary disk and thus avoids this problem. Based on the results, we determined detection limits for protoplanetary mass of $1.45\pm 0.04$ M$_{\rm J}$ and $2.29 ^{+0.03}_{-0.04}$ M$_{\rm J}$ for the gaps at 25 and 95 AU, respectively, in the TW Hya disk. 
These limits agree with the results of previous exoplanet explorations, though modeling of disk observations points to significantly lower upper limits of order 0.1 M$_{\rm J}$.
The latest ExAO instruments have a magnitude limit of $R\leq$12, but many YSOs are fainter than this. Our results indicate that searching for very young exoplanets with classical AO instruments can be an option.

\acknowledgments
The authors wish to thank Jim Lyke and Mihoko Konishi for helping set up OSIRIS's new pipeline after April 2016. We also express our gratitude to Katsuhiro Murata and Patrick Woudt for performing the IRSF/SIRIUS observations. We would like to thank the anonymous referees for their constructive comments and suggestions that improved the quality of the paper.
The data presented herein were obtained at the W. M. Keck Observatory, which is operated as a scientific partnership between the California Institute of Technology, the University of California, and the National Aeronautics and Space Administration. Use of the observatory was made possible by the generous financial support of the W. M. Keck Foundation.
The observations were carried out within the framework of the Subaru-Keck time exchange program, with travel expenses supported  by the National Astronomical Observatory of Japan, which operates the Subaru Telescope. This research made use of the SIMBAD database and the VizieR catalog access tool, both provided by CDS, Strasbourg, France. The original description of the VizieR service was published in A\&AS 143, 23. This publication makes use of data products from the Two Micron All Sky Survey, which is a joint project between the University of Massachusetts and the Infrared Processing and Analysis Center/California Institute of Technology, and is funded by the National Aeronautics and Space Administration and the National Science Foundation.

T.U. is supported by a Japan Society for Promotion of Science (JSPS) Fellowship for Research, and this work was partially supported by the Grant-in-Aid for JSPS Fellows (Grant Number 17J00934). M.T. is partly supported by the JSPS Grant-in-Aid (No. 15H02063).

The authors wish to acknowledge the very significant cultural role and reverence that the summit of Mauna Kea has always had within the indigenous Hawaiian community.  We are most fortunate to have the opportunity to conduct observations from this mountain.

\bibliographystyle{apj}                                                               
\bibliography{library}                                                                

\appendix 
\section{Theoretical Simulation of Hydrogen Emissions within the Circumstellar Disk}

\begin{table}[h] \centering
  \caption{Adopted Model Parameters of TW Hya}
  \begin{tabular}{cc}\\ \hline\hline
  Parameters & Value \\ \hline
  Pre-shock gas temperature [${\rm K}$] & 100 \\
  Pre-shock gas density [$\rm g/{cm}^3$] & $\rho_{\rm CSD}$ ($a$) \\
  Planet Radius [R$_{\rm J}$] & 2 \\
  Dust/gas ratio & 0.0 \\
  Magnetic field [$\rm G$]& 0.0 \\
  Metal line optical depth [${\rm cm}^{-1}$]& 0.0 \\
  \hline
  \end{tabular}
  \label{model parameter}
  \tablecomments{
  $\rho_{CSD}$ is the gas density of circumstellar disk where $a$ is the planet's semi-major axis} 
\end{table}

Model parameters for mass constraint are listed in Table \ref{model parameter}.
We assume that the gas temperature before the shock surface is 100 K. Before the strong shock, the gas temperature depends on the primary star emission and bow shock, which hardly makes the gas thermal energy larger than the shock energy of the strong shock \citep[e.g.][]{Tanigawa2012}.
However, this parameter hardly affects structures of flow and radiation field after the shock unless it becomes over thousands of K, because shock energy is at least $\sim$3$\times10^4$ K in hydrogen-line-emitting region \citep{Aoyama2017P}.

The gas density of pre-shock region is assumed to be as same as that of the circumstellar disk based on the 3D hydrodynamic simulation of \cite{Tanigawa2012}.
Planet disturbance on global disk structure, e.g. global gap structure, is not included.

Hydrogen lines are strongly emitted when the pre-shock velocity $v_0 \gtrsim$ 30 km/s. In a slower case, post-shock hydrogen cannot be excited enough to emit the strong hydrogen lines.
This lower limit of the velocity corresponds to the free fall velocity on the planet surface of $\sim$0.5 M$_{\rm J}$.
The planet radius affects the surface integral of 1D numerical simulation results especially near the lower limit mass \citep{Aoyama2017P}. 
However, this hardly affect the hydrogen line luminosity because the outer region in the circumplanetary disk with larger area contributes more to the luminosity than the planet surface or inner region in the circumplanetary disk with smaller area does.

We also assume dust, magnetic field, and metal line cooling are negligible in hydrogen line emitting region around protoplanets.
Since dust settling to midplane and the gas accreting near the planet come from high altitude of the circumstellar disk, hydrogen-line-emitting gas flow hardly contains dust.
When the pre-shock gas is ionized, the magnetic field changes the shock structure and lowers the post-shock temperature \citep[e.g.][]{Draine1980}.
In post-shock region, magnetic pressure can affect the post-shock structure \citep{Hollenbach&McKee1979}.
When the optical depth of hydrogen lines becomes thick enough not to cool the gas effectively, the metal line cooling becomes important \citep{Hollenbach&McKee1989}.

\end{document}